%% file: paper.tex
\documentclass[sigconf]{acmart}

\newcommand{\ignore}[1]{}
\usepackage{fancyhdr}
\usepackage[normalem]{ulem}

\newcommand{\code}[1]{\texttt{#1}}

\usepackage{listings}
\usepackage{color}
\usepackage{algpseudocode}
\usepackage{blindtext}
\usepackage{color}   
\usepackage{amssymb,amsmath}
\usepackage{comment}
\usepackage{enumerate}
\usepackage{framed}
\usepackage{array}      
\usepackage{balance}
\usepackage{xspace}
\usepackage{multirow}
\usepackage{booktabs}
\usepackage{amsmath}
\usepackage{datetime}
\usepackage{subcaption}
\usepackage{graphicx}
\usepackage{microtype}

\graphicspath{{figures/}{pics/}}
\DeclareGraphicsExtensions{.pdf,.jpeg,.png}

\title{Stochastic Synthesis for Stochastic Computing}
\author{Vincent T. Lee, Armin Alaghi, Luis Ceze, Mark Oskin \\ University of Washington}

\input{00-abstract}

\renewcommand\footnotetextcopyrightpermission[1]{} 
\begin{document}
\sloppy
\maketitle
\pagestyle{plain}


\input{01-introduction}
\input{02-background}
\input{03-approach}
\input{04-methodology}
\input{05-evaluation}
\input{06-limitations}

\input{07-related-work}
\input{08-conclusion}



\balance
\bibliographystyle{ACM-Reference-Format}
\bibliography{references}


\end{document}

%% file: 00-abstract.tex
\begin{abstract}

Stochastic computing (SC) is an emerging computing technique which offers higher computational density, and lower power over binary-encoded (BE) computation.
Unlike BE computation, SC encodes values as probabilistic bitstreams which makes designing new circuits unintuitive.
Existing techniques for synthesizing SC circuits are limited to specific classes of functions such as polynomial evaluation or constant scaling.
In this paper, we propose using stochastic synthesis, which is originally a program synthesis technique, to automate the task of synthesizing new SC circuits.
Our results show stochastic synthesis is more general than past techniques and can synthesize manually designed SC circuits as well as new ones such as an approximate square root unit.

\end{abstract}

%% file: 01-introduction.tex
\section{Introduction}

\noindent Stochastic computing (SC) is an emerging computation technique which has enjoyed renewed interest as a promising paradigm for low power, dense, and error resilient computation.
Unlike binary-encoded (BE) values, SC values are encoded in unary bitstreams (stream of 1s and 0s).
The value of a bitstream is enumerated by summing the number of 1s and 0s and dividing by the bitstream length.
This encoding allows arithmetic operations to be implemented with simple gates.
For example, multiplication in SC is implemented by a single two-input AND gate (Fig. ~\ref{fig:sc_circuits}a); given a bitstream $X = 11101110$ $(p_X = 0.75)$ and $Y = 01110010$ $(p_Y = 0.5)$, the bitwise AND yields the product $Z = 01100010$ $(p_Z = 0.375)$.
What SC circuits gain in density and lower power, they lose in terms of run time since bitstreams take multiple cycles to execute whereas equivalent BE circuits take only a handful of cycles to execute.

Unlike BE logic which have well-known formulations for transforming a target function to CMOS logic, values in SC are encoded temporally and present additional challenges such as correlation between bitstreams which evade human intuition.
A key challenge with SC is that aside from the well-known set of SC circuits, designing new SC circuits is a manual process that requires significant design effort, and theoretical insight.
Furthermore, existing techniques are limited to specific classes of functions.
For instance, prior work has shown how to synthesize SC circuits for functions like polynomial evaluation~\cite{strauss, qian08}, rational functions~\cite{saraf16}, and constant scaling~\cite{ting-dfts17}.
Previous synthesis methods for sequential SC circuits such as~\cite{saraf16} are limited in that they cannot synthesize sequential SC circuits that exploit correlation.
This leaves a large space of circuits that defy existing SC analysis techniques outside the known set of synthesizable function classes.

In this paper, we propose leveraging stochastic synthesis~\cite{stoke} for automatically synthesizing SC circuits from test case specifications.
Note that stochastic synthesis and stochastic computing (SC) are unrelated techniques.
Stochastic synthesis is a program synthesis technique which treats the space of programs as a high dimensional space, where each program has a user-defined cost.
The synthesizer then traverses the space of programs by iteratively sampling ``close by'' programs and moving towards programs with better cost.
Transitions in the space are generated by a set of program rewrites which transform one program into another.
Unlike previous work in SC, the stochastic synthesis technique proposed in this paper is not limited to a specific class of functions or circuits; it can also provide approximate SC circuits if an exact solution cannot be found or is not known to exist.

\vspace{1mm}
\noindent Our contributions are as follows:
\begin{itemize}
\item A formulation of stochastic synthesis for automatically synthesizing SC circuits.
\item Showing stochastic synthesis can generate approximate SC circuits in lieu of exact solutions.
\item A novel approximate square root circuit which highlights the efficacy of our synthesis technique.
\end{itemize}

The rest of the paper is organized as follows.
Section~\ref{sec:background} provides background for SC and stochastic synthesis.
Section~\ref{sec:approach} outlines our synthesis formulation.
Section~\ref{sec:methodology} outlines our evaluation methodology and Section~\ref{sec:results} summarizes our results.
Section~\ref{sec:limitations} discusses limitations and Section~\ref{sec:related-work} compares our technique against prior work.

%% file: 02-background.tex
\section{Background}
\label{sec:background}

\noindent This section provides background on stochastic computing and stochastic synthesis which are distinct orthogonal concepts.

\subsection{Stochastic Computing}

{
  \setlength{\belowcaptionskip}{0pt}
  \setlength{\abovecaptionskip}{0pt}
 \begin{figure}[b]
   \centering
  \includegraphics[width=\linewidth]{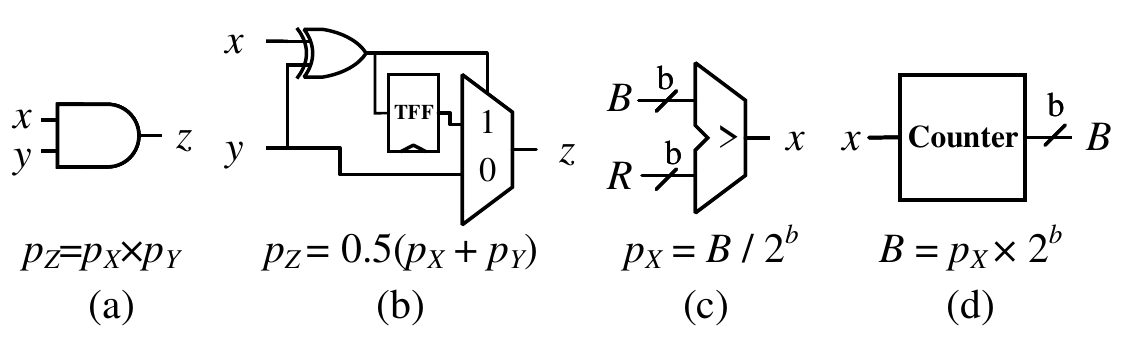}
  \caption{SC circuits: (a) multiplier, (b), scaled adder, (c) digital-to-stochastic converter, and (d) stochastic-to-digital converter.}
  \label{fig:sc_circuits}
\end{figure}
}

\noindent Stochastic computing (SC) is a technique that dates back to the 1960s~\cite{gaines69} and proposes using unary bitstreams (stream of 0s and 1s) to encode numbers as probabilities.
Bitstreams in SC are often referred to as \textit{stochastic numbers} (SNs).
The encoded \textit{value} of a SN is defined as the sum over each position in the SN divided by the SN length $N$.
Since each bit in a SN has uniform weight, the maximum unique values an SN with length $N$ can represent is $N+1$; in other words, a SN with length $N$ has a precision of $log_2(N)$.

SNs typically use either unipolar or bipolar representations.
In unipolar representations, 1s in the SN are ascribed a weight of $+$1 and 0s are ascribed a weight of 0. 
For instance, the unipolar SN $X = 01000011$ encodes the value 0.375 since there are three 1s and the SN length $N = 8$.
In bipolar representations, 1s are weighted as +1 while 0s are weighted as $-$1.
This allows bipolar representations to encode negative values (i.e. the range [$-$1, 1]).
For instance, the same SN $X = 01000011$ encodes the value $-$0.250 since there are three 1s, five 0s, and the SN length is 8.

The unary encoding allows for arithmetic operations in SC to be realized with a small number of gates.
Fig.~\ref{fig:sc_circuits} shows implementations of SC multiplication and addition for unipolar representations.
A unique constraint in SC is that the output precision is forced to equal the input precision of any SC operation.
As a result, arithmetic operations where the required output precision is higher than the input precision suffer from quantization errors.
For instance, SC addition can either only be implemented as a scaled addition $f(p_X, p_Y) = 0.5(p_X + p_Y)$ or as a saturating add $f(p_X, p_Y) = \code{min}(1.0, p_X + p_Y)$.

To generate SNs from BE values, we typically use digital-to-stochastic (D/S) converters (Fig.~\ref{fig:sc_circuits}c) which are also known as stochastic number generators (SNGs).
An SNG takes a BE value $x$ and a random number, and compares them; this generates the desired series of 1s and 0s since the probability of the comparator generating a 1 is proportional to the BE value over the length of the SN.
The choice of random number generator that drives the SNG is critical as it governs the correlation between SNs.
Two SNs generated with the same RNG are positively correlated while two SNs generated by different RNGs will be uncorrelated.
Many SC arithmetic operations require either correlated or uncorrelated input SNs to operate correctly.
To convert from the stochastic domain to BE values, we use a stochastic-to-digital (S/D) converter (Fig.~\ref{fig:sc_circuits}d) which is realized using a counter.

\subsection{Stochastic Synthesis}

{
  \setlength{\abovecaptionskip}{0pt}
\begin{figure}[b]
  \begin{algorithmic}[1]
    \Procedure{Synthesis}{$I$, $\beta$, $C(X)$}
    \State $P \gets$ random program of length $I$
    \State $B \gets P$ // Initialize best program
    \While{compute budget not exhausted}
    \State $R \gets $ random rewrite rule
    \State $P' \gets R(P)$ // Generate proposal program
    \State $\alpha \gets \code{min}(1, \code{exp}(-\beta (C(P') - C(P))))$ // Evaluate cost
    \If{$random\_number(0, 1) < \alpha$}
    \State $P \gets P'$ // Accept proposal
    \Else
    \State pass // Reject proposal
    \EndIf
    \If{$C(P') < C(B)$}
    \State $B \gets P'$ //Update best program
    \EndIf
    \EndWhile
    \State return $B$
    \EndProcedure
  \end{algorithmic}
  \caption{High level stochastic synthesis algorithm.}
  \label{fig:synthesis}
\end{figure}
}

\noindent Stochastic synthesis is a program synthesis technique used for superoptimization of program binaries and compiling to idiosyncratic instruction sets~\cite{stoke, stoke-fp, scaling-up-superopt}.
Stochastic synthesis is an instance of Markov chain Monte Carlo where the space of programs is treated as a high-dimensional space.
Each program $P$ is ascribed a cost calculated by a user-defined function $C(P)$ which captures correctness and/or optimality.
The synthesizer then iteratively traverses the space of programs towards lower cost programs, similar to gradient descent algorithms.
Intuitively, this effectively samples promising sectors of the program space since exhaustive enumeration is prohibitively expensive.

A summary of the stochastic synthesis algorithm is shown in Fig.~\ref{fig:synthesis} for target program of length $I$.
The initial program in the stochastic search is randomly generated.
From this initial program, the search iteratively generates proposals for better programs by randomly applying one of several rewrite rules $R(P)$ (discussed in Section~\ref{sec:approach}).
All decisions when executing rewrite rules are performed randomly.
For example, when applying the replace operand rewrite rule the synthesizer must first select an input operand to overwrite and a new operand to replace it with.
In this case, the deleted register is randomly selected from the set of existing instruction operands and the new register is randomly selected from the pool of available registers.

The set of rewrite rules available to the synthesizer must be ergodic which guarantees that given infinite resources the search will eventually explore all possible programs.
Given a current program $P$ and a proposed candidate program $P' = R(P)$, the search either accepts or rejects the candidate program.
If a program proposal is accepted, the current program becomes the proposed program; if the candidate is rejected, the current program remains the same.
The candidate generation process is then repeated until an optimal program is found or the computational budget is expended.

The probability of a proposed program being accepted or rejected is based on its cost $C(P')$ relative to the cost of the current program $C(P)$ and is computed using the Metropolis ratio:
\[ \alpha(P, P') = \code{min}(1, \code{exp}(-\beta (C(P')-C(P))) \]
\noindent Intuitively, this probability distribution forces the search to always accept a proposal with better cost while allowing the search to still accept less optimal programs.
Accepting less optimal, higher cost programs during the search is crucial for enabling the search to escape local minima in the program space.
The value for $\beta$ is tuned experimentally, similar to how the learning rate is tuned for machine learning applications.

SC circuits are an ideal candidate for stochastic synthesis for several reasons.
First, many known SC circuits use a handful of gates, which limits the search space.
Stochastic synthesis is notorious for poor scalability with increasing program size so confining the search to small programs significantly improves the chance of success.
Second, precision or SN length in SC does not affect circuit functionality; this allows the same SC circuit synthesized with one SN length to generalize to longer SNs.
On the other hand, binary-encoded computation require different circuits to process different precision values.
Third, relative to large software instruction sets, the number of hardware primitives is small, which reduces the search space significantly.
All together, these considerations significantly reduce the search space when compared to software formulations of stochastic synthesis which have larger and more complex program spaces.

%% file: 03-approach.tex
\section{Synthesis Formulation}
\label{sec:approach}

\noindent This section outlines the synthesis formulation we use to design stochastic circuits.
At a high level, our synthesizer takes a target function specification and a set of test case inputs.
More important, these test cases express the correlation conditions between input SNs under which the circuit must operate correctly.
The goal of the synthesizer is then to find a circuit implementation that best approximates the target function given the specified input test cases.

\subsection{Instruction Set and Program Definitions}

\noindent Existing software formulations of stochastic synthesis target software instruction sets like x86 and are agnostic to the notion of cycle count or time.
Hardware design on the other hand must incorporate the notion of cycle count or time into the formulation to expose the semantics of state elements.
The hardware instruction set and semantics for our stochastic synthesis formulation are shown in Table~\ref{tab:isa}.
Our instruction set is reminiscent of the primitives provided by structural Verilog, and includes primitive gates (ex. AND, OR, XOR) in addition to well-known primitives for SC (ex. T-flip flop (TFF), multiplexor (MUX)).
Each instruction is composed of an opcode indicating its operation, input operands, and output operands.

Unlike software formulations of stochastic synthesis, a hardware program is considered invalid if the same destination register is assigned multiple times (doubly driven wire) or the program forms a combinational loop.
In our formulation, we express a circuit as a program of hardware instructions.
Programs that realize invalid circuits are ascribed maximal cost ($C(P) = 1.0$) to discourage the search from these areas.
To prevent doubly driven or undriven wires, we impose single static assignment over the program and prevent rewrite rules from overriding assigned destination registers.
We also require the user to specify the target program length.
Finally, it is important to note that the spatial nature of hardware makes programs agnostic to instruction order.

{
  \captionsetup[table]{skip=0pt}
\begin{table}[t]
  \centering
  \caption{Hardware program instruction set.}
  \label{tab:isa}
  \begin{tabular}{@{}ll@{}} \toprule
    Instruction & Semantics \\ \midrule
    AND src, trg, dst & dst[n] $\leftarrow$ src[n] \& trg[n] \\
    OR src, trg, dst & dst[n] $\leftarrow$ src[n] $\vert$ trg[n] \\
    XOR src, trg, dst & dst[n] $\leftarrow$ src[n] $\oplus$ trg[n] \\
    NOT src, dst & dst[n] $\leftarrow$ $\neg$ src[n] \\
    PASS src, dst & dst[n] $\leftarrow$ src[n] \\
    DFF src, dst & dst[n] $\leftarrow$ src[n-1] \\
    TFF src, dst & dst[n] $\leftarrow$ dst[n-1] $\oplus$ src[n-1] \\
    MUX src, trg, sel, dst & dst[n] $\leftarrow$ src[n] if sel[n] else trg[n] \\ \bottomrule
  \end{tabular}
\end{table}
}

\subsection{Specification and Cost Function}

\noindent The input specification to our synthesizer is a set of test cases and their target output values.
A test case is defined as a set of input bindings to input operands and desired output SN value; the user must specify the number of input operands which can be derived from the target function to synthesize.
For a given program $P$, we define the cost of a program as the average absolute error between the expected output SN value and the result SN value produced by the circuit.
The result SN value is calculated by simulating the circuit for each set of input bindings.
We define the total cost $C(P)$ of the program as the average absolute error over all test cases; this ensures that the cost function is agnostic to SN length and test case count which reduces how often we need to tune $\beta$.

To generate test cases, we select from LFSR, Van der Corput, or Halton (base = 3) sequences for generating input SN operands.
test case selection and coverage directly impacts the cost function and ultimately the behavior of the synthesized circuit.
test case selection can also be manipulated to express the conditions under which the desired SC circuit will operate.
For instance, test cases can be intentionally generated with correlated or uncorrelated inputs to tell the synthesis process to find a circuit that operates correctly with correlated or uncorrelated operands respectively.

If the optimal operating conditions are unknown, the user can ask the synthesizer to determine what the optimal correlation should be; to do this, the user supplies duplicate operands to the synthesizer and lets the synthesizer determine which ones to use.
For example, if we were to try and synthesize a SC subtractor but did not know whether the input operands needed to be correlated or uncorrelated, we would supply three input SNs $X$, $X'$, and $Y$, where $X$ and $X'$ have the same value.
test cases would be generated such that $X$ and $Y$ are uncorrelated and $X'$ and $Y$ are positively correlated.
The synthesizer will then figure out whether the uncorrelated or correlated inputs are unnecessary and will leave one disconnected if necessary in the synthesized result.
The key drawback of this technique is that it increases the search space of potential programs by introducing additional input operands.

\subsection{Program Generation and Rewrite Rules}

\noindent Candidate programs are generated by randomly selecting from a set of rewrite rules.
The set of rewrite rules is typically a combination of rules which locally perturb the program and rules which impose more global modifications.
Each rewrite rule is assigned a selection probability which governs how often it is used in the search.
We generally find that ascribing higher selection probability to local rewrite rules and lower probabilities to more global rewrite rules works well.
Intuitively, this allows the search to spend sufficient time exploring local minimum before moving on to another local minimum.

The set of rewrite rules used by our stochastic synthesizer is shown in Table~\ref{tab:rewrite_rules}.
Unlike traditional software program synthesis, we do not have a swap instruction rewrite rule since hardware is agnostic to program order.
We also add a swap all operands rewrite rule which randomly selects two operands $ra$ and $rb$, and replaces every instance of $ra$ with $rb$ and $rb$ with $ra$.
This effectively swaps the connectivity of two wires in the circuit netlist without requiring the synthesizer to traverse potentially many high cost intermediary circuits.
Finally, we also employ random restarts~\cite{random-restart} which is a technique that reinitializes the search to a random program.
This effectively forces the synthesizer to explore a different portion of the program space and potentially rescues it from getting stuck in deep local minima in the search space.

{
  \captionsetup[table]{skip=0pt}
\begin{table*}[t]
  \centering
  \caption{Program rewrite rules and selection probabilities. Minor rewrite rules like operand replacement have much higher selection probability than major rewrite rules like random restart.}
  \label{tab:rewrite_rules}
  \begin{tabular}[t]{@{}lll@{}} \toprule
    Rewrite Rule & Prob. & Description \\ \midrule
    Replace Operand & 0.817 & Replace a random instruction's input register with new random operand. \\
    Replace Opcode  & 0.091 & Replace a random instruction's opcode with a new opcode of the same arity. \\
    Replace Instruction & 0.045 & \begin{tabular}{@{}l@{}}Replace the entirety of a random instruction with a new randomly generated instruction. \end{tabular} \\
    Swap All Operands & 0.045 & \begin{tabular}[t]{@{}l@{}} Randomly selects two registers $ra$ and $rb$. \\ Replaces every instance of $ra$ with $rb$ and $rb$ with $ra$ in the program. \end{tabular} \\
    Random Restart & 0.001 & Replaces the entire program with a new random program. \\
    \bottomrule
  \end{tabular}
\end{table*}
}

{
  \captionsetup[table]{skip=0pt}
\begin{table*}[t]
  \centering
  \caption{Summary of synthesis benchmarks. Our stochastic synthesizer can synthesize existing SC arithmetic circuits as well as approximations for new ones.}
  \label{tab:benchmarks}
  \begin{tabular}{@{}cccccccc@{}} \toprule

    Arithmetic Unit &
    \begin{tabular}{@{}c@{}}Function\\$f(p_X, p_Y)$\end{tabular} &
    \begin{tabular}{@{}c@{}} Baseline\\Design \end{tabular} &
    \begin{tabular}{@{}c@{}}Baseline\\Length\end{tabular} &
    \begin{tabular}{@{}c@{}} Synthesized\\Length\end{tabular} &
      \begin{tabular}{@{}c@{}} Absolute\\Error\end{tabular} &
        Correct \tabularnewline \midrule

         Scaled Adder       &  $\frac{(p_X + p_Y)}{2}$ & ~\cite{lee17-date}     &  3                    &  3 &  0.027 &  Yes \tabularnewline
         Subtractor         &  $|p_X-p_Y|$ &  ~\cite{sc-correlation} &  1                         &  1 &  0 &  Yes \tabularnewline
     Uncorrelated Multiplier      &  $p_Xp_Y$ &  ~\cite{gaines69}       &  1                            &  2 &  0.021 &  Yes \tabularnewline
     Division            &  $p_X/p_Y$ &  ~\cite{cordiv}         &  2                           &  2 &  0.038 &  Yes \tabularnewline
     Scale $\times$1/4        &  $0.25p_X$ &  ~\cite{ting-dfts17}    &  4                           &  5 &  0 &  Yes \tabularnewline
     Scale $\times$1/3        &  $0.33p_X$ &  ~\cite{ting-dfts17}    &  4                           &  5 &  0 &  Yes \tabularnewline
     Scale $\times$1/2        &  $0.5p_X$ &  ~\cite{ting-dfts17}    &  2                            &   2 &  0 &  Yes \tabularnewline
     Scaled ReLU         &  $max(0.5, p_X)$ &  ~\cite{sc-relu}         &  11                   &  16 &  0 &  Yes \tabularnewline
     Correlated Multiplier &  $p_Xp_Y$ & N/A                      &  N/A                         &  4 &  0.035 &  N/A \tabularnewline
     Square Root         &  $\sqrt{p_X}$ &  ~\cite{gaines69}                        &  N/A                      &  5 &  0.024 &  N/A \tabularnewline
     Sine                &  $\frac{sin({2\pi}p_X)+1}{2}$ &  ~\cite{gaines69}                        &  N/A                 &  8 &  0.068 &  N/A \tabularnewline
     Exponentiation         &  $p_X^{p_Y}$ & N/A                        &  N/A                       &  7 &  0.031 &  N/A \tabularnewline
     Cosine              &  $\frac{cos({2\pi}p_X)+1)}{2}$ &  ~\cite{gaines69}                        &  N/A                 &  10 &  0.15 &  N/A  \tabularnewline
    \bottomrule
  \end{tabular}
\end{table*}
}

In our experiments, we experimentally find that the parameter $\beta = 2$ in the Metropolis ratio produces good results.
Since we ascribe a maximum cost of 1.0 to invalid circuit, $\beta = 2$ allows invalid circuits to be accepted with small but non-negligible probability.
We still want the search to explore such invalid programs since it is often necessary to traverse invalid programs to reach new valid ones.
Finally, to improve cost evaluation performance our synthesizer performs combinational loop checks and dead code elimination.
Programs that are invalid are not evaluated which saves computational resources.

%% file: 04-methodology.tex
\section{Synthesis Benchmarks} \label{sec:methodology}

\noindent We evaluate our stochastic synthesis formulation for both well-known arithmetic operations in addition to operations which have known but inefficient implementations.
In our evaluation, we will limit the set of synthesis benchmarks to unipolar SC circuits, but we expect our synthesis formulation to generalize to bipolar and other SC representations.
Table~\ref{tab:benchmarks} shows the set of synthesis benchmarks we use to evaluate the capabilities of stochastic synthesis.
We deem a synthesized circuit as correct if it logically equivalent, or has the same cost or better than known solutions in prior work.

We are particularly interested in benchmarks which involve state elements and require feedback since such solutions are more likely to defeat human design intuition and existing synthesis techniques.
For each benchmark, we execute the synthesizer for at least 1 million proposals, which takes less than 10 minutes, and return the solution with best cost.
We increase the initial instruction count and number of proposals as needed for operations that have previously unknown solutions.
Note that the number of evaluated proposals corresponds directly to the compute budget in Fig.~\ref{fig:synthesis}.
If the synthesizer encounters an exact solution (cost of zero/no error), it immediately terminates the search and returns that solution.

A key strength of stochastic synthesis is that even if the synthesizer fails to reach an exact solution (no error), it will still return a solution that approximates the target function.
This makes stochastic synthesis an ideal solution for attempting to synthesize SC implementations of functions which elude known SC synthesis techniques and currently have inefficient solutions.
Examples of such functions include square root and exponentiation, and trigonometric functions such as sine, and cosine which can be implemented using Adaptive Digital Elements (ADDIE)~\cite{gaines69}.
However, ADDIE circuits are expensive and inefficient because they often require counters and additional auxiliary inputs.
For our synthesis benchmarks, we modify the target sine and cosine functions such that their range is in the unipolar domain (i.e. [0,1]).

%% file: 05-evaluation.tex
\section{Evaluation Results}
\label{sec:results}

\noindent This section presents synthesis results and quantifies the efficacy of our synthesis formulation.

{
  \captionsetup[table]{skip=0pt}
\begin{table*}[t]
  \centering
  \caption{Synthesis results for uncorrelated multiplication, scaled addition, constant scaling by 0.25, constant scaling by 0.33, square root, exponentiation, correlated multiplication, and polynomial evaluation.}
  \label{tab:synthesis_results}
  \begin{tabular}{@{}m{0.1\linewidth}m{0.2\linewidth}m{0.2\linewidth}m{0.2\linewidth}m{0.2\linewidth}} \toprule
    \centering Operation & \centering Uncorrelated Multiplier & \centering Scaled Addition & \centering Scale by 1/4 & \centering Scale by 1/3 \tabularnewline \midrule
    \centering Known Solution &
    \includegraphics[width=\linewidth]{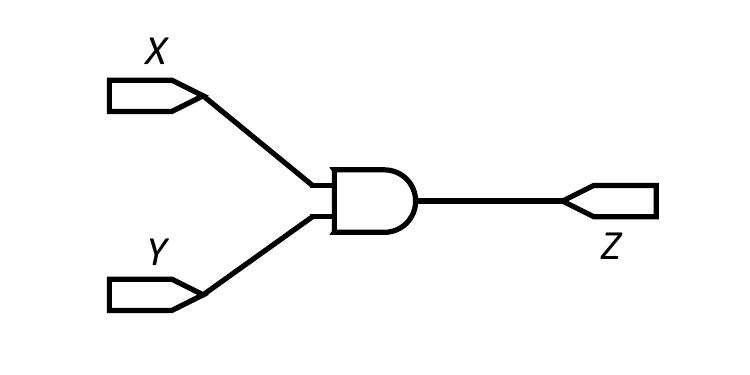} &
    \includegraphics[width=\linewidth]{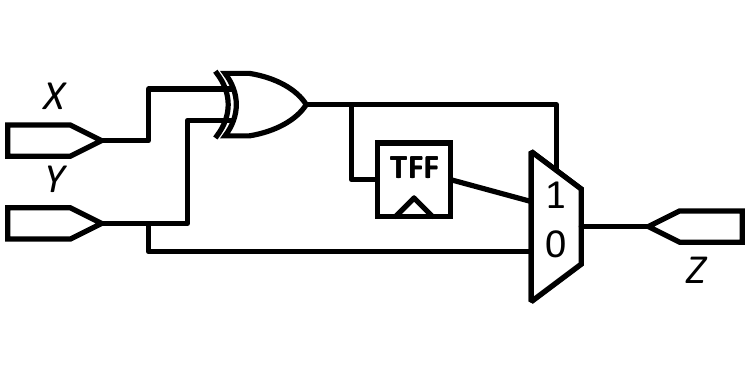} &
    \includegraphics[width=\linewidth]{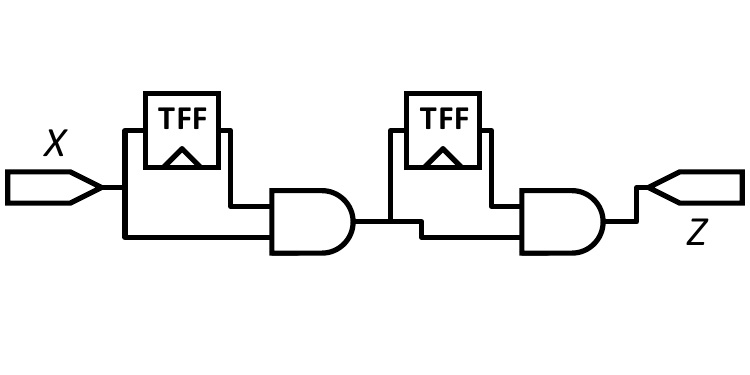} &
    \includegraphics[width=\linewidth]{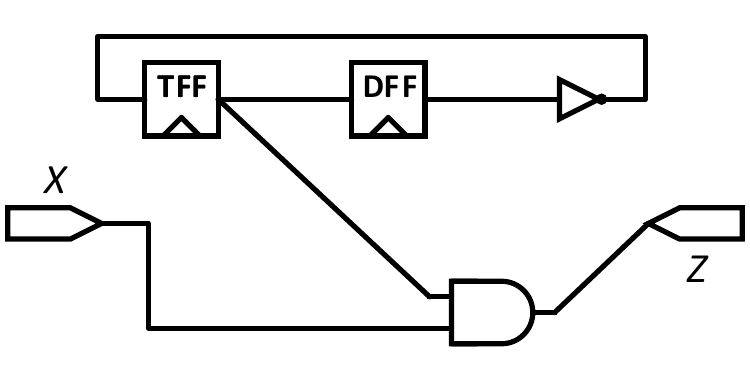} \tabularnewline

    \centering Synthesized Solution &
    \includegraphics[width=\linewidth]{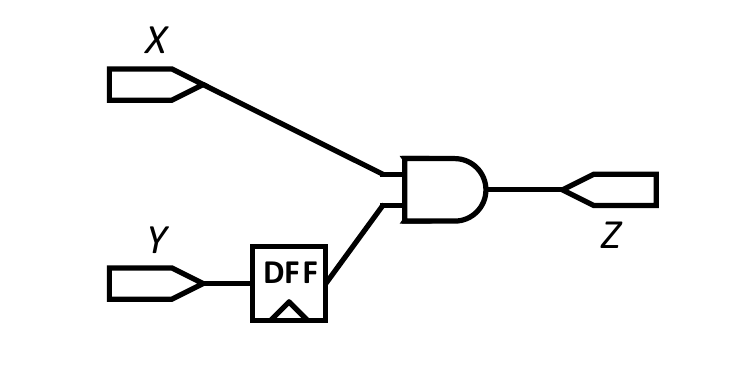} &
    \includegraphics[width=\linewidth]{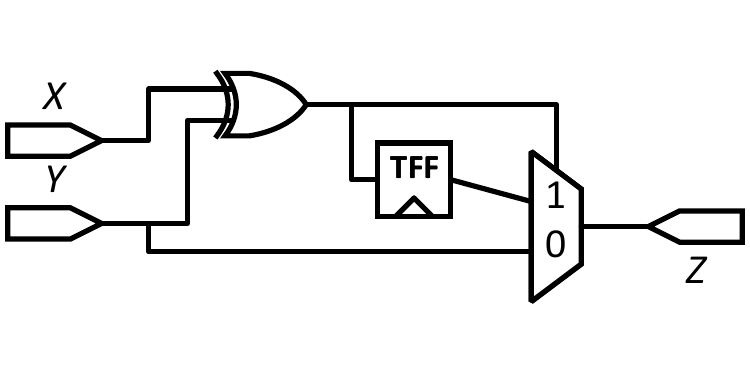} &
    \includegraphics[width=\linewidth]{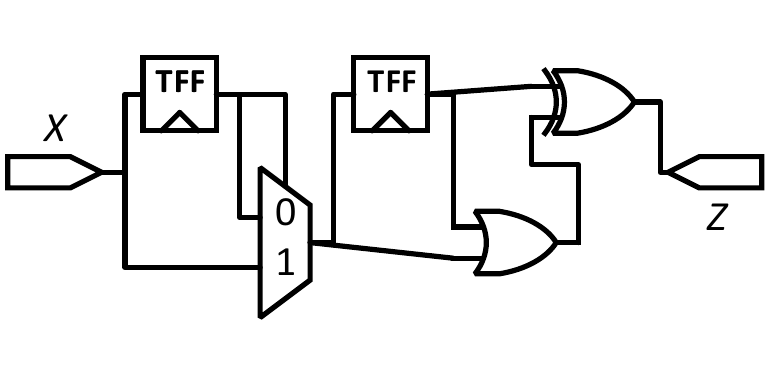} &
    \includegraphics[width=\linewidth]{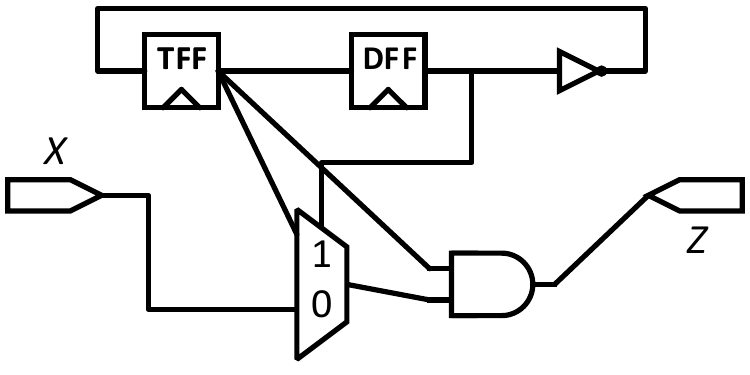} \tabularnewline

    \toprule

    \centering Operation & \centering Correlated Multiplier & \centering Square Root & \centering Exponentiation & \centering Divider \tabularnewline
    \midrule
    \centering Known Solution &
    \centering No Known Solution &
    \centering \includegraphics[width=\linewidth]{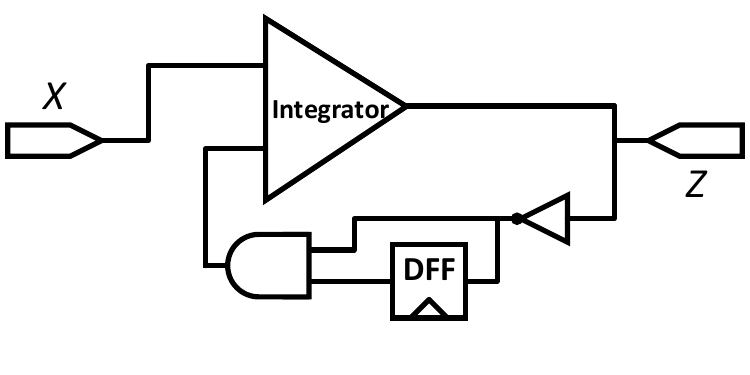} &
    \centering No Known Solution &
    \includegraphics[width=\linewidth]{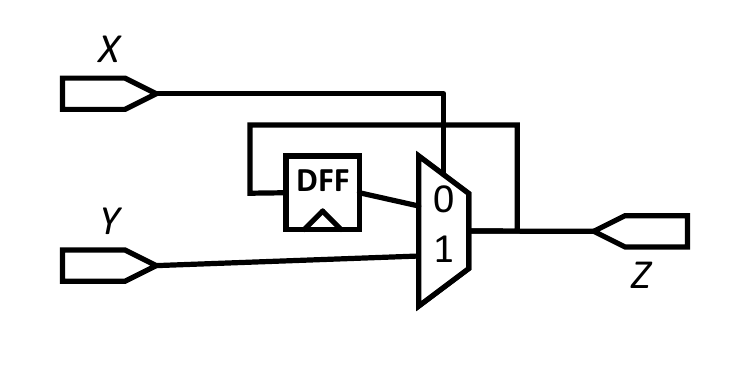} \tabularnewline
    \centering Synthesized Solution &
    \includegraphics[width=\linewidth]{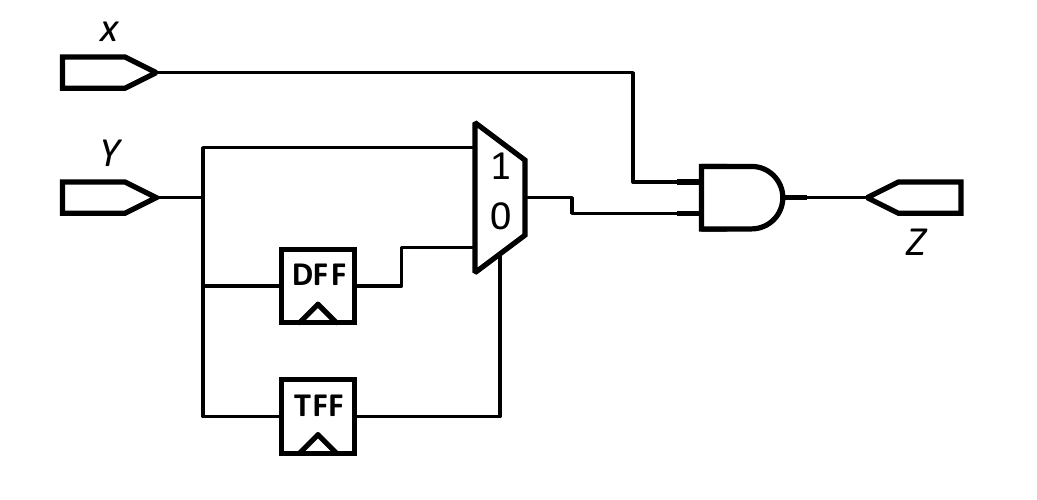} &
    \includegraphics[width=\linewidth]{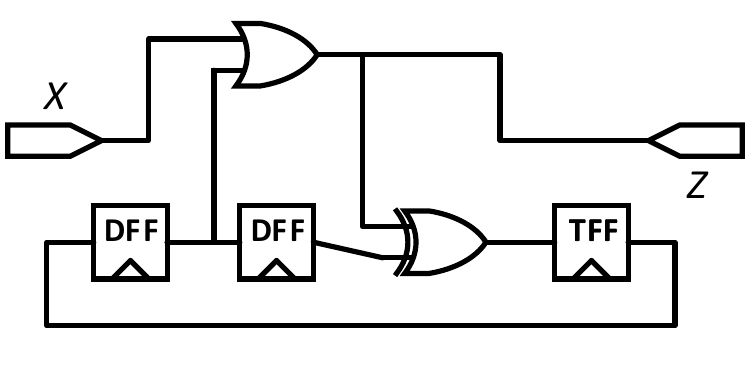} &
    \includegraphics[width=\linewidth]{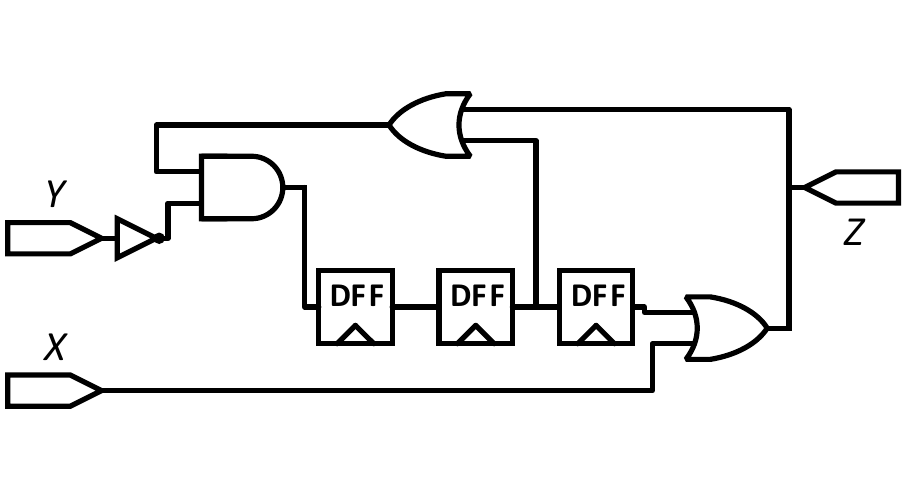} &
    \includegraphics[width=\linewidth]{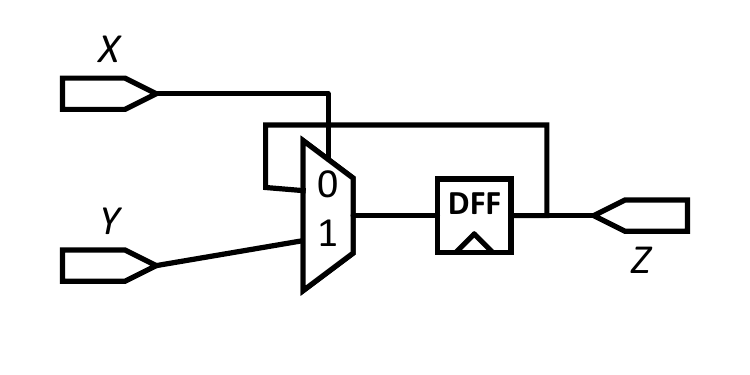}
    \tabularnewline
    \bottomrule
  \end{tabular}
  \setlength{\textfloatsep}{0.1mm}
\end{table*}
}

\subsection{Synthesis Results}

\noindent In general, our stochastic synthesizer is able to quickly synthesize correct implementations of known SC circuits such as the subtractor, uncorrelated multiplier, and scaled adder.
Note, that existing tools are incapable of designing the subtractor and correlation agnostic adder which exploit or manage correlation.
On average, we find the synthesizer evaluates about 2000 program proposals per second.
Interestingly, we find that for uncorrelated multiply benchmark the synthesizer finds a solution that correctly includes an isolator (DFF) before the AND gate.
Similar experiments indicate that the synthesizer is able to perform isolator insertion as it tries to improve the circuit.
We also attempt to synthesize a SC multiplier to handle cases where input SNs are correlated (the original SC multiplier assumes uncorrelated SNs).
As shown in Table~\ref{tab:synthesis_results}, our synthesizer is able to find a circuit which uses a T-flip flop, D-flip flop, and multiplexor to break the correlation between the two input SNs before feeding the SNs into an AND gate to perform the multiplication.
For scaled addition, our synthesizer is able to correctly identify the correlation agnostic adder proposed in~\cite{lee17-date}.

We also find that the synthesizer can discover correct implementations (Table~\ref{tab:synthesis_results}) of scaling by constant circuits equivalent to those generated by CEASE~\cite{ting-dfts17}.
For these particular benchmarks the synthesizer produces solutions which are suboptimal in terms of number of resources (cost function does not optimize for circuit size) but are logically equivalent to correct implementations.
For instance, the synthesized solution for scale by 1/4 can reduce the multiplexor to an AND gate, and the solution for the scale by 1/3 can optimize away the multiplexor.
Fortunately, standard logic reduction techniques reduce such synthesized solutions to their smaller, more optimal implementations.
Interestingly, the synthesizer finds the correlation agnostic variant for the scale by 1/4 benchmark and the correlation sensitive variant for scale by 1/3 benchmark~\cite{ting-dfts17}.
More importantly, this result shows that our synthesis formulation can discover circuits with both sequential elements and feedback paths which is not possible with techniques in prior works.

Our synthesizer is also able to find approximate implementations of functions that are difficult to analyze and manually design solutions for.
For instance, we are able to synthesize a new approximate square root and exponentiation ($f(x, y) = x^y$) circuit as shown in Table~\ref{tab:synthesis_results}.
Our solutions do not require expensive integrators, and auxiliary SNGs used in previous work using ADDIE circuits~\cite{gaines69}.
In addition, the synthesized square root and exponentiation circuits shown in Table~\ref{tab:synthesis_results} cannot be synthesized by existing SC synthesis techniques.
Interestingly, we find that these synthesized solutions uses a set of sequential elements reminiscent of the modulo counters or LFSR shift register architecture used in scaling by constant circuits.
Unlike the scaling by constant circuits or LFSRs, the feedback loop also takes $p_X$ and $p_Y$ as input.

\begin{figure}[t]
  \centering
  \begin{lstlisting}

  \end{lstlisting}
  \begin{subfigure}{0.23\textwidth}
    \centering
    \includegraphics[width=\linewidth]{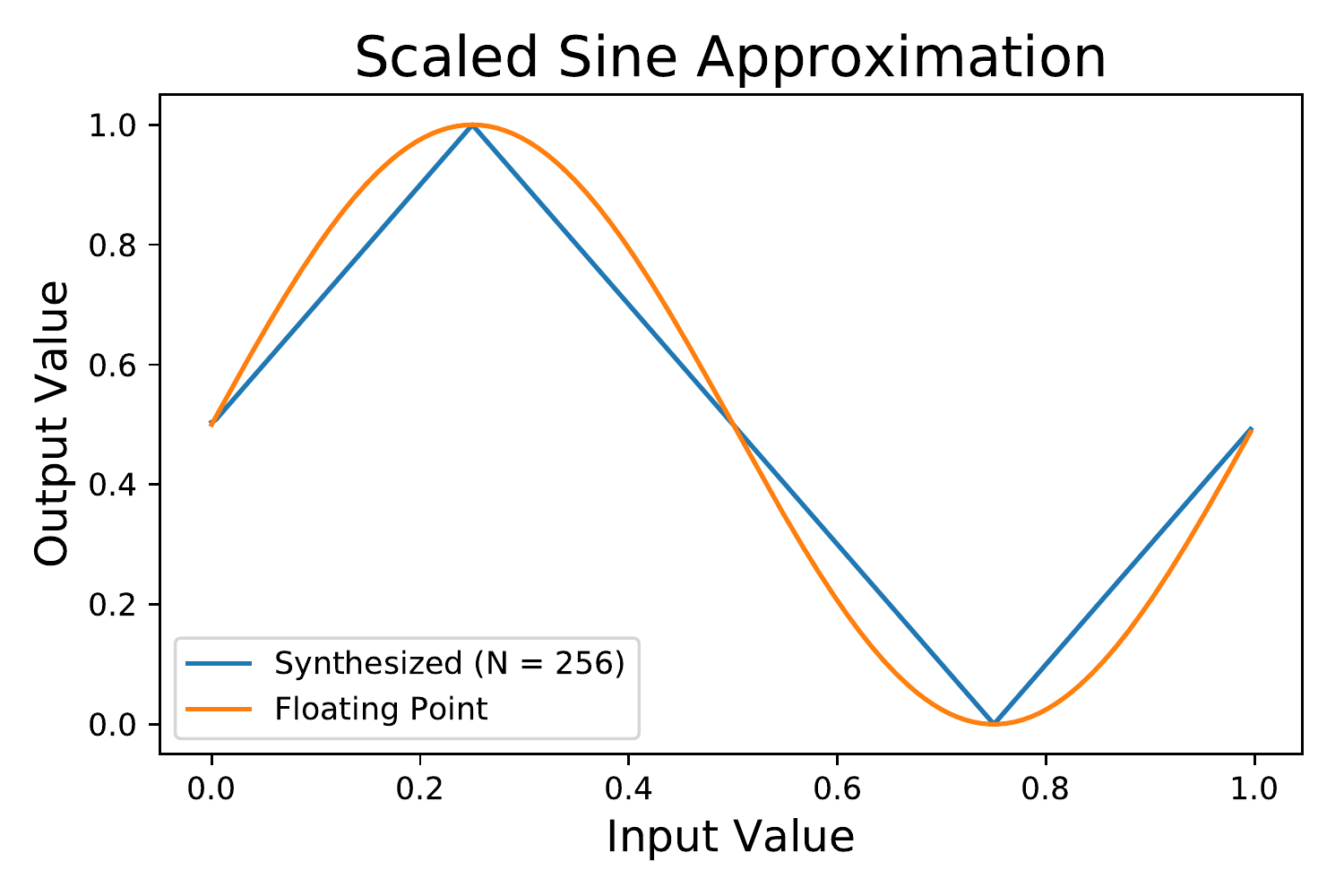}
    \caption{Scaled sine approximation}
  \end{subfigure}%
  \begin{subfigure}{0.23\textwidth}
    \centering
    \includegraphics[width=\linewidth]{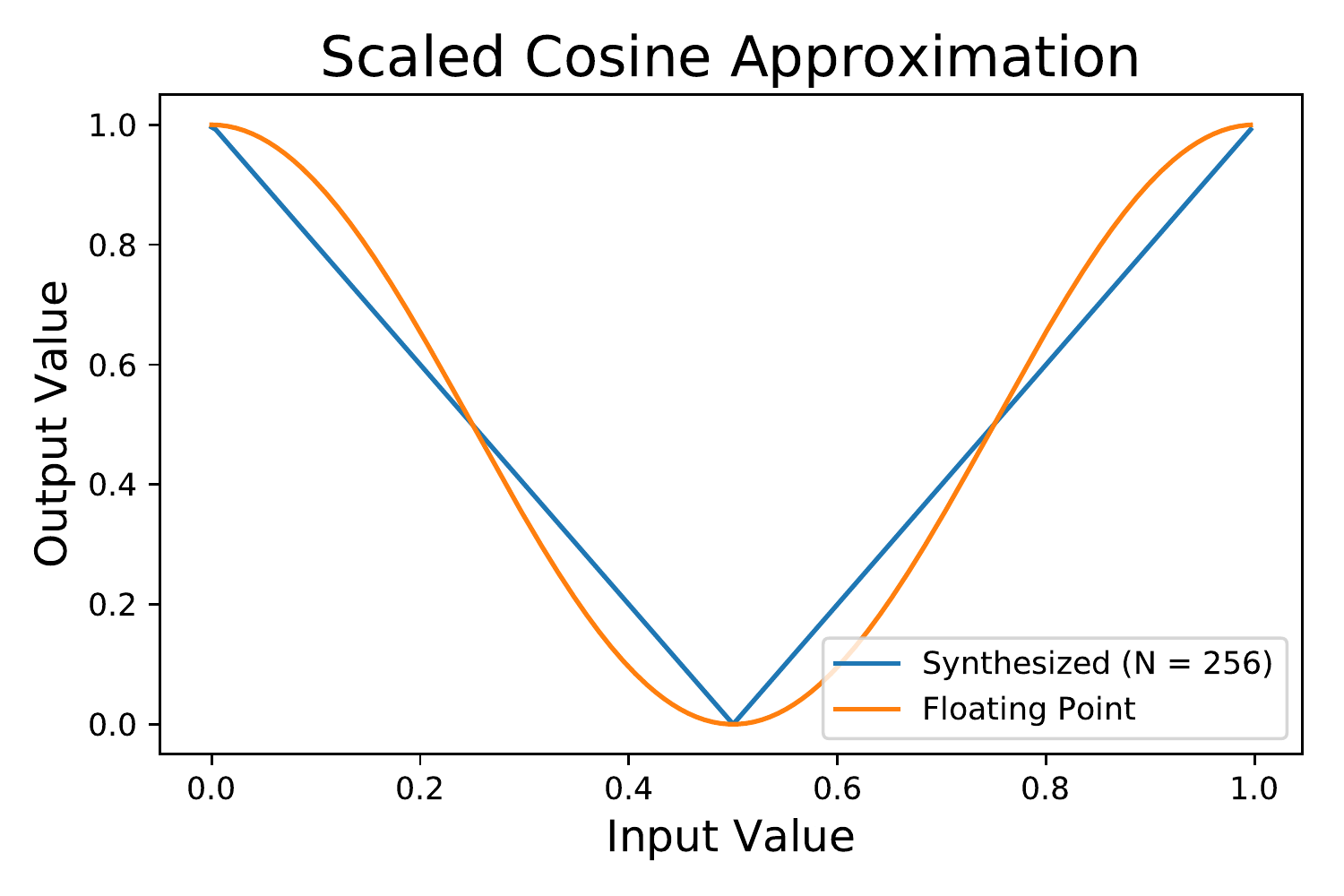}
    \caption{Scaled cosine approximation}
  \end{subfigure}
  \caption{Synthesized sawtooth approximations of scaled sine and cosine functions using Van der Corput inputs (N=256).}
  \label{fig:scaled_trig}
\end{figure}

For other more difficult benchmarks like scaled sine and cosine, we find the synthesizer is not able to find as optimal solutions.
The synthesizer is still able to find a reasonable sawtooth wave approximation using inputs generated by Van der Corput sequences (Fig.~\ref{fig:scaled_trig}, but these solutions are not ideal.
Interestingly, the synthesized sine and cosine solutions reduce to finite state machines which qualitatively appear similar to the counter-based rectified linear unit (ReLU) proposed in~\cite{sc-relu}.
The failure of the synthesis formulation to find a good solution does not preclude the existence of a better solution nor guarantees a better solution exists and alludes to some of its limitations discussed later.

For larger circuits, we find that instantiating the program with more instructions than the known solution size improves the search result.
These extra instructions serve as extra degrees of freedom and many are often deleted during dead code elimination since they do not drive any part of the circuit.
But by increasing the program size, it allows the search to find larger but logically equivalent variations of the known solution; this increases the number of potential solutions and hence the number of optimal local minima in the program space.

\subsection{Generality of Synthesized Circuits}

\noindent We find that synthesized circuits generalize to arbitrary SN length and validates the assumption that it is sufficient to synthesize a general SC solution using a fixed SN length.
An example of synthesized circuits that generalize to arbitrary SN length are the constant scaling circuits.
In this case, the synthesizer finds both the modulo counter-based implementation and correlation agnostic implementations which generalize to arbitrary SN length.

Another instance of this precision generalization is our synthesized approximate square root circuit.
Fig.~\ref{fig:square_root_error} compares the actual result generated by the synthesized square root circuit for several SN lengths, and compares them against the expected floating point function; as our results show, the synthesized circuit behavior remains the same across all SN lengths.
The circuit was synthesized using SNs using Van der Corput sequences but also works (albeit with modest errors) for SNs generated with LFSR and Halton (base = 3) sequences.

{
  \setlength{\abovecaptionskip}{0pt}
\begin{figure}[b]
  \centering
  \includegraphics[width=\linewidth]{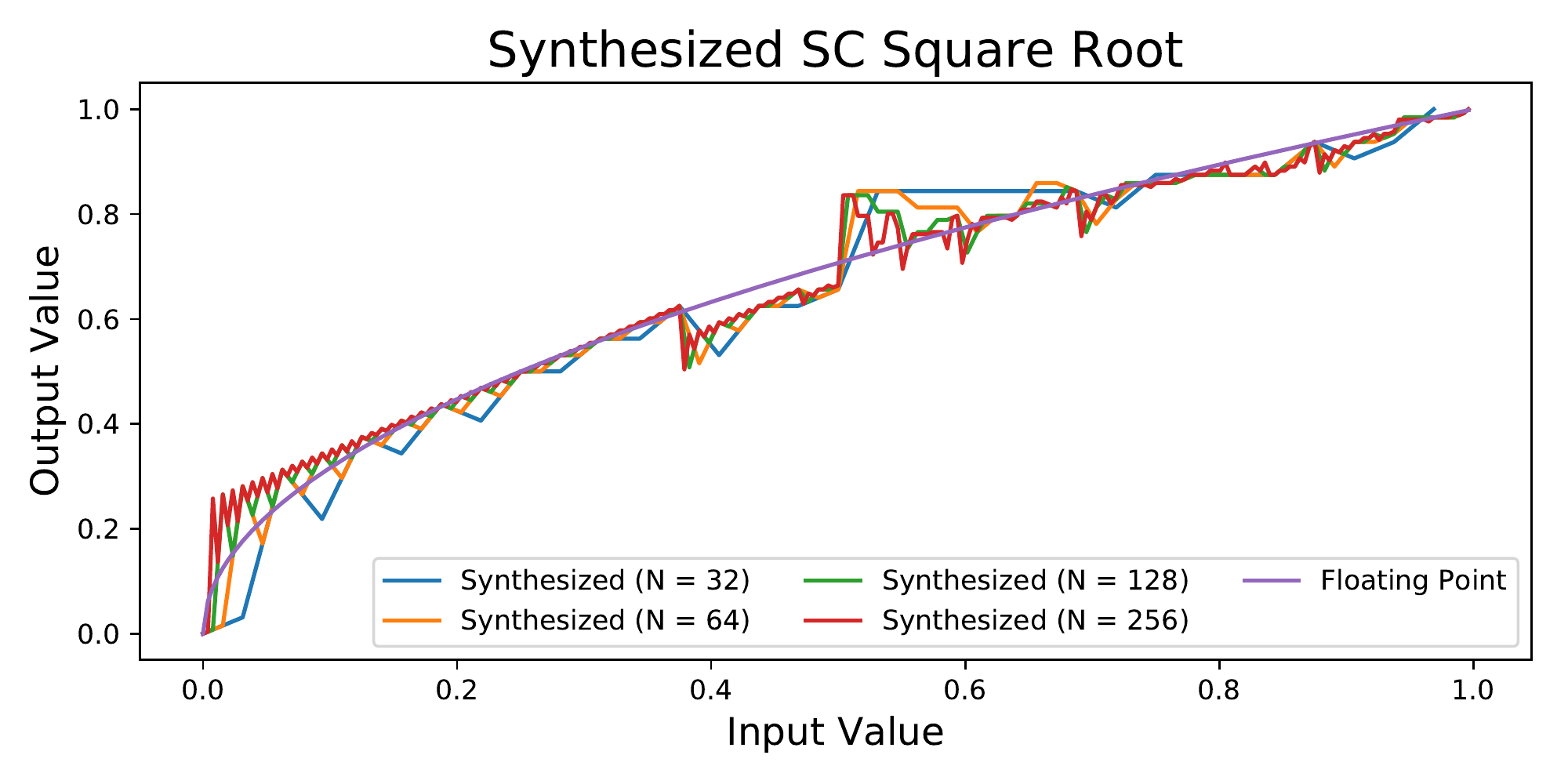}
  \caption{Synthesized approximate SC square root function. The synthesized circuit generalizes to different SN lengths.}
  \label{fig:square_root_error}
\end{figure}
}

\subsection{Stochastic Synthesis versus Exhaustive Search}

A simpler alternative to stochastic synthesis is brute force exhaustive search of all possible circuits and guarantees the optimality of the solution.
Recall, the ideal stochastic circuit is small otherwise risks losing its power, density, and energy efficiency advantages over BE circuits.
To compare the efficiency of exhaustive search against stochastic synthesis, we calculate the total number of possible circuits the exhaustive search must evaluate.

Our results find that exhaustive search is only a viable alternative to stochastic synthesis for circuits of up to 2 gates ($1.64 \times 10^4$ possible circuits).
Each additional instruction in the program increases the search space by roughly three to four orders of magnitude which makes this brute force enumeration quickly intractable.
For instance, a circuit of size 5 has $3.73 \times 10^{13}$ possible candidates, many of which are invalid due to combinational loops and unconnected wires.
As a result, many of the circuits shown in Table~\ref{tab:synthesis_results} would be beyond the capabilities of exhaustive search.

Finally, exhaustive search is complementary and not orthogonal to stochastic synthesis~\cite{scaling-up-superopt}.
Exhaustive search guarantees the optimality of the identified solution but does not scale beyond very small circuits.
Stochastic synthesis on the other hand is more scalable but does not guarantee optimality.
When combined, the stochastic synthesis can efficiently explore the space of large programs while exhaustive search can ensure that no better solutions exist within the set of small circuits.

%% file: 06-limitations.tex
\section{Scalability and Extensions}\label{sec:limitations}

\noindent Stochastic synthesis is clearly an excellent match for SC since circuits tend to be small in size in order to retain their advantage over BE.
However, stochastic synthesis has its own unique strengths and limitations.
In particular, stochastic synthesis suffers from the scalability issues that impede most program synthesis techniques.
Recall that the program space increases exponentially with target program size and number of operands.
Consequently, stochastic synthesis requires exponentially more computation resources to explore the larger space which limits scalability to modest sized circuits.
Less general SC synthesis methods such as STRAUSS scale more reasonably since there is a direct relationship between the specification and the implemented solution.
Finally, the failure of the stochastic synthesizer to return an exact solution neither guarantees an exact solution exists, nor does it guarantee a better approximation exists.

The stochastic synthesis formulation presented in this paper was only evaluated for unipolar SN representations but can easily be extended to other representations or cost functions.
For instance, our synthesis formulation can target bipolar or inverted bipolar representations by adjusting the input test cases and cost function so that they interpret the bitstreams properly.
Our synthesis formulation can also be extended to optimize for other cost metrics such as circuit size or even SN correlation (again by modifying the cost function).
In fact, the stochastic synthesis cost function can be modified to reflect any desired cost metric as the optimization target.
Finally, the stochastic synthesis can also be used to optimize existing circuits instead of synthesizing circuits from scratch (also known as stochastic superoptimization).
This amounts to bootstrapping the synthesizer to a part of the design space which is known to at least be locally optimal.

%% file: 07-related-work.tex
\section{Related Work}

\label{sec:related-work}

\noindent A key strength of stochastic synthesis over previous techniques is that it is not limited to any particular class of functions or circuit properties.
As a result, it is an excellent tool for identifying circuits which may be unintuitive for a designer or for generating approximate circuits for circuits which have no known solution.
Tools such as STRAUSS~\cite{strauss} and ReSC~\cite{qian08} are limited to synthesizing combinational SC circuits without sequential elements for polynomial evaluation.
To use polynomial evaluation to approximate functions, the user must identify and tune the parameters of the desired polynomial beforehand.
In many cases, the desired polynomial is not obvious when trying to approximate functions.

Similarly, SC synthesis techniques for sequential SC circuits are limited to rational functions and cannot identify ways to manage correlation.
Stochastic synthesis does not have these limitations and can synthesize both polynomial target functions and correlation manipulating sequential circuits without any prior information.
The key trade off for this generality is that stochastic synthesis is limited to synthesizing small gates.
However, this limitation is acceptable since the ideal stochastic circuits should be small or risk losing their density, power, and energy efficiency advantage over BE circuits.

Stochastic synthesis is also able to correctly synthesize circuits for scaling by constants and reproduce the results originally shown by Ting et al.~\cite{ting-dfts17}.
Our results show that stochastic synthesis is able to both identify the modulo counter-based solutions as well as the correlation agnostic solutions.
We also showed that stochastic synthesis can also discover when it is appropriate to insert isolators (ex. uncorrelated multiplier) or identify ways to break correlation (ex. correlated multiplier).
Furthermore, stochastic synthesis is able to find approximate implementations for a target function when an exact solution may not exist which makes it more powerful than existing SC synthesis methods.

Finally, stochastic synthesis is a test case-driven technique which is different from analytical SC circuit synthesis techniques which assume Bernoulli random variable inputs.
Using a test case driven technique allows the designer to precisely specify the input SNs the circuit will see.
These test cases can be used to also encode the correlation conditions between SNs which analytical models used in previous work do not consider.
As shown by our results, our synthesis technique can identify circuits with sequential elements with feedback paths that exploit SN correlation or autocorrelation properties.
Examples of circuits include the square root and exponentiation circuits which cannot be identified using existing analytical SC synthesis techniques.

%% file: 08-conclusion.tex
\section{Conclusions}
\label{sec:conclusions}

\noindent This paper proposes stochastic synthesis for designing SC circuits, and presents a formulation for synthesizing SC circuits from test case inputs.
We show that our synthesis formulation can synthesize logically equivalent solutions to manually designed SC circuits in prior work.
We also show our synthesis technique is able to find approximate solutions for functions like square root when exact solutions are not found or may not exist.
We find synthesized solutions generalize to different SN lengths and show that, unlike previous SC synthesis techniques, stochastic synthesis is not limited to any particular class of functions or circuits.